\documentclass[10pt,conference]{IEEEtran}
\usepackage{cite}
\usepackage{amsmath,amssymb,amsfonts}
\usepackage{algorithmic}
\usepackage{graphicx}
\usepackage{textcomp}
\usepackage{xcolor}
\usepackage[hyphens]{url}
\usepackage{fancyhdr}
\usepackage[colorlinks=true,linkcolor=black,anchorcolor=black,citecolor=blue,filecolor=black,menucolor=black,runcolor=black,urlcolor=blue]{hyperref}
\usepackage{multirow}
\usepackage{amsmath}
\usepackage{algorithmic}
\usepackage{graphicx}
\usepackage{textcomp}
\usepackage{xcolor}

\usepackage{graphicx}
\usepackage{textcomp}
\usepackage{xcolor}
\usepackage{enumitem}
\usepackage[numbers,sort&compress]{natbib}

\usepackage{subcaption}
\usepackage{graphicx}
\usepackage{textcomp}
\usepackage{xcolor}
\usepackage{xargs}
\usepackage{xspace}
\usepackage{braket}
\usepackage{bbold}
\usepackage{tikz}
\usepackage{caption}
\usepackage{bm}
\usepackage{subcaption}
    
\newcommand{\hmma}{\texttt{HMMA}\xspace}
\newcommand{\fp}{\texttt{FP}\xspace}
\newcommand{\half}{\texttt{FP16}\xspace}
\newcommand{\bft}{\texttt{BF16}\xspace}
\newcommand{\tf}{\texttt{TF32}\xspace}
\newcommand{\ours}{\textit{MPGemmFI}\xspace}

\usepackage{fancyhdr}
\fancypagestyle{firstpage}{
  \fancyhf{}

  \fancyhead[C]{\normalsize{SC 2023 Submission. Confidential Draft: DO NOT DISTRIBUTE}} 
  \fancyfoot[C]{\thepage}
}

\newcommand{\hpcayear}{2024}

\newcommand{\hpcasubmissionnumber}{830}

\begin{document}

\title{MPGemmFI: A Fault Injection Technique for Mixed Precision GEMM in ML Applications}

\newcommand{\hpcapubid}{0000--0000/00\$00.00}
\newcommand\hpcaauthors{First Author$\dagger$ and Second Author$\ddagger$}
\newcommand\hpcaaffiliation{First Affiliation$\dagger$, Second Affiliation$\ddagger$}
\newcommand\hpcaemail{Email(s)}

\author{
\IEEEauthorblockN{Bo Fang\IEEEauthorrefmark{1}, Xinyi Li\IEEEauthorrefmark{2},
Harvey Dam\IEEEauthorrefmark{2},
Cheng Tan\IEEEauthorrefmark{7},
Siva Kumar Sastry Hari\IEEEauthorrefmark{3},
Timothy Tsai\IEEEauthorrefmark{3},
Ignacio Laguna\IEEEauthorrefmark{4},\\
Dingwen Tao\IEEEauthorrefmark{5},
Ganesh Gopalakrishnan\IEEEauthorrefmark{2},
Prashant Nair\IEEEauthorrefmark{6},
Kevin Barker\IEEEauthorrefmark{1},
Ang Li\IEEEauthorrefmark{1}
}
\IEEEauthorblockA{\IEEEauthorrefmark{1} Pacific Northwest National Laboratory \{bo.fang, ang.li, kevin.barker\}@pnnl.gov}
\IEEEauthorblockA{\IEEEauthorrefmark{6} University of British Columbia, \{prashant.nair\}@ece.ubc.ca}
\IEEEauthorblockA{\IEEEauthorrefmark{2} University of Utah 
Email: \{xin\_yi.li, harvey.dam,u0028245\}@gatech.edu}
\IEEEauthorblockA{\IEEEauthorrefmark{3} NVIDIA 
Email: \{shari,timothyt\}@nvidia.com}
\IEEEauthorblockA{\IEEEauthorrefmark{4} Lawrence Livermore National Laboratory 
Email: \{ilaguna\}@llnl.gov}
\IEEEauthorblockA{\IEEEauthorrefmark{5} Indiana University 
Email: \{ditao\}@iu.edu}
\IEEEauthorblockA{\IEEEauthorrefmark{7} Google 
Email: \{chengtan\}@google.com}
}


\maketitle 

\begin{abstract}
Emerging deep learning workloads urgently need fast general matrix multiplication (GEMM). To meet such demand, one of the critical features of machine-learning-specific accelerators such as NVIDIA Tensor Cores, AMD Matrix Cores, and Google TPUs is the support of mixed-precision enabled GEMM. For DNN models, lower-precision \fp data formats and computation offer acceptable correctness but significant performance, area, and memory footprint improvement. While promising, the mixed-precision computation on error resilience remains unexplored. To this end, we develop a fault injection framework that systematically injects fault into the mixed-precision computation results. We investigate how the faults affect the accuracy of machine learning applications 
Based on the error resilience characteristics, we offer lightweight error detection and correction solutions that significantly improve the overall model accuracy if the models experience hardware faults. The solutions can be efficiently integrated into the accelerator's pipelines.
\end{abstract}

\section{Introduction}
\label{sec:intro}

The reliability of large-scale systems faces a significant challenge with the rise of transient hardware faults. This issue is exacerbated by shrinking feature sizes, increasing system scales, and power constraints~\cite{201410, jack, facebook, 2014franck}. Graphics Processing Units (GPUs) serve as key accelerators in High-Performance Computing (HPC) systems, and their extensive deployment makes them especially vulnerable to transient hardware faults~\cite{largescale_gpu, raditation-induced, extremescaleresilience}. Although error correction code (ECC) is commonly used to protect memory, caches, and register files in GPUs, a substantial portion of GPU chips comprising computation units remain without ECC protection. Consequently, applications accelerated by GPUs may suffer from silent data corruptions (SDCs) when transient hardware faults propagate through the GPUs' datapaths. Addressing the SDC challenge is crucial to ensure the dependable operation of large-scale systems that use GPUs as accelerators.

The performance of emerging machine-learning applications heavily relies on efficiently executing General Matrix Multiplication (GEMM). For instance, deep neural network (DNN) models containing fully-connected and convolution layers spend more than 90\% of their execution time on GEMMs when running on GPUs~\cite{jia2014learning}. Similarly, a substantial portion (60\% in single precision and 45\% in mixed precision) of the execution time for BERT models is dedicated to GEMM computations~\cite{bert-gemm}. To accelerate GEMM, specialized GPU architectures have been designed with two common features: (i) a hardware-optimized computational pipeline for matrix-based operations and (ii) native support for mixed-precision floating-point (\fp) computation, combining low-precision and high-precision formats in the fused multiply-and-add operations, the core computation in GEMM. However, despite sharing similar technology features with traditional GPUs, these specialized architectures are susceptible to hardware faults, and it is essential to understand the impact of such faults on mixed-precision GEMM computation, which dominates the computation cycles. This understanding will enable the adoption of advanced fault tolerance techniques to enhance the error resilience of machine learning models.

While several prior studies have explored the resilience of machine learning models, \ours stands out for its specific focus on the target system. For instance, Zhang et al.~\cite{cluster1} concentrated on understanding the consequences of memory errors, while Rojas et al.~\cite{cluster2} introduced errors that modify model checkpoints. Additionally, PyTorchFI~\cite{pytorchfi} and TensorFI~\cite{tensorfi} provided interfaces for injecting faults into the output tensor of model layers in PyTorch and TensorFlow, respectively. In a recent work~\cite{mphpc}, a compiler-based fault injection technique was used to characterize the impact of hardware faults on mixed-precision HPC applications, revealing that combining single and mixed \fp formats led to comparatively more frequent and severe silent data errors than double-precision \fp. However, these cutting-edge techniques fall short in handling the intricacies of GEMM instructions with mixed-precision formats, making them incapable of systematically mimicking the errors that impact mixed-precision computations. Our research addresses this gap by systematically presenting a novel approach to introduce representative faults. This helps advance the understanding of the resilience of GEMM accelerators in mixed-precision scenarios.

This study aims to design a methodology that enables the development of precision-aware fault tolerance techniques for GEMM-specialized GPUs. To this end, our methodology must accomplish three key tasks: Firstly, the methodology should be capable of characterizing how a transient hardware fault affects the error resilience of an application conducting mixed-precision GEMM computation. The methodology must be able to introduce faults systematically during GEMM operations. Secondly, the characterization study should be conducted in the context of real-world applications. For applications with inherent error masking capabilities, such as deep learning models~\cite{tensorfi,gpli} and iterative solvers~\cite{bv,2014franck}, the methodology must explore the gap between the error resilience of a mixed-precision MM operation and the end-to-end result of an application. Lastly, the observation and insights obtained should guide the design of a lightweight precision-aware fault tolerance approach that specifically identifies and corrects errors given a mixed-precision \fp computation pattern. 


To the best of our knowledge, our work is the first to characterize the error resilience of mixed-precision GEMM computation and devise specific fault tolerance techniques. Our work aims to enhance the reliability of applications accelerated by specialized GPUs. We propose \ours, a fault injection framework to systematically introduce bit-flip faults in the result of mixed-precision GEMM operations. We conduct a comprehensive fault injection study on representative deep learning models. Utilizing findings from three lower-precision \fp formats, we explore the design space for safeguarding machine learning tasks from hardware faults based on their numerical features. Unlike existing methods, our techniques do not necessitate pre-training or profiling to obtain a model's data distribution, and we make no assumptions about the details of the ML models. 

\vspace{0.05in}
\noindent \textbf{Contributions}: We make the following four key contributions:
\begin{itemize}[topsep=0pt,leftmargin=*]
\renewcommand{\labelitemi}{\scriptsize$\blacksquare$}
    \item We present a fault injection framework \ours to systematically inject faults into the mixed-precision GEMM that uses three precision formats (\half, \bft, and \tf) on NVIDIA Tensor Core, one of the most representative specialized GPU architectures.
    \item We leverage \ours to conduct large-scale fault injection campaigns on five representative deep neural network (DNN) models to estimate their error resilience against different \fp formats.  Our analysis reveals the key features of the three \fp formats and shows that compared to \half and \tf, \bft is more vulnerable to errors by 3.97x and 3.11x than \half and \tf on average.
    \item We propose three exponent-only error detection and correction mechanisms that automatically detect and correct the result of a \bft multiplication without \textit{a priori} knowledge about the ML application data. The proposed techniques significantly improve the model accuracy when affected by hardware faults by, on average 75\%. 
    \item We extrapolate the cost of implementing these mechanisms within the hardware pipeline and show that our design only requires a limited number of logic operations.
\end{itemize}



\vspace{0.05in}
\noindent \textbf{Lessons learned:} This paper unveils the error resilience of machine learning models utilizing GEMMs with different \fp formats (\bft, \half, and \tf). While the fundamental behaviors of FP numeric computations remain consistent across these formats, the varying proportions of exponent versus mantissa bits lead to distinct outcomes against faults. Our evaluation demonstrates that the \bft format exhibits relatively lower resilience to hardware faults, owing to its larger number of exponent bits and the ratio of the exponent-to-mantissa bits. This insight guides our proposition of lightweight error detection and correction techniques based solely on exponents without duplicating \fp operations within GEMM accelerators.

In the remainder of this paper, we first introduce the features of modern mixed-precision GPU architectures, using NVIDIA Tensor Cores as an example (Section~\ref{sec:back}). Next, we detail our systematic approach to characterize the error resilience of mixed-precision GEMMs on NVIDIA Tensor Cores in Section~\ref{sec:method}. We present the error resilience characteristics of five representative DNN models utilizing mixed-precision FP computation in Section~\ref{sec:eval}. To enhance the error resilience of specific mixed-precision formats, we propose three lightweight error detection and correction mechanisms in Section~\ref{sec:detection_correction}. We demonstrate the benefits of applying these techniques and estimate their overall cost. Finally, we conclude in Section~\ref{sec:conclusion}.
\section{Background}
\label{sec:back}


This section provides an overview of the different precision formats, aside from \texttt{FP32} and \texttt{FP64}. We highlight the key features of NVIDIA's Tensor Core accelerators, focusing on their role in mixed-precision GEMM operations. Additionally, we introduce \ours and NVBitFI to emphasize the capabilities of the former in this context.

\subsection{Diverse floating-point formats}

\begin{figure}[h!]
    \vspace{-0.05in}
    \centering
    \includegraphics[width=0.7\linewidth]{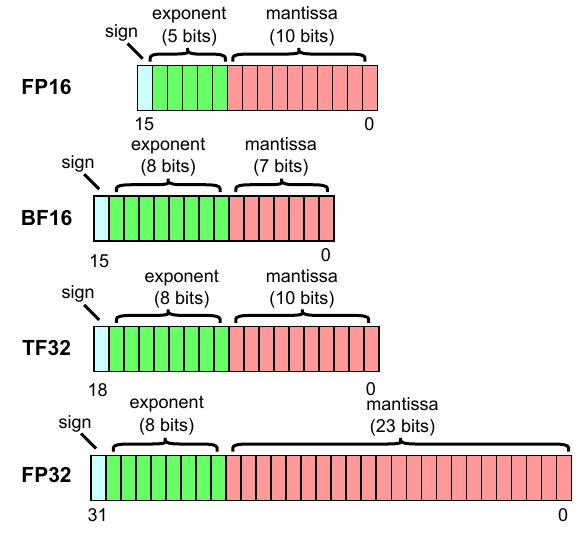}
    \caption{Floating-point formats of \half, \bft and \tf, compared to \texttt{FP32}. The boundaries between the exponent and mantissa bits are aligned across different formats.}
    \label{fig:fp}
    \vspace{-0.05in}
\end{figure}

We explore three floating-point formats: half (\half)\cite{754}, Brain Floating Point Format (\bft)\cite{bf16}, and TensorFloat-32 (\tf)\cite{tf32}. Figure\ref{fig:fp} illustrates their respective floating-point formats, including the numbers of sign, exponent, and mantissa bits used. Generally, lower-precision formats compared to \texttt{FP32} offer faster computation and a reduced memory footprint, where the drawback of lower precision is often not critical for deep learning workloads. For instance, \half uses half the memory of \texttt{FP32} while maintaining consistent training accuracy~\cite{mixed-precision}. \bft and \tf are designed to optimize the \half format for deep learning applications. Both formats expand the dynamic range of \half to match that of \texttt{FP32}, with \tf providing even more mantissa bits for increased precision requirements.

When a bit flip occurs in a \fp number, the consequences of the fault will be influenced by the characteristics of the bits-to-\fp number conversion system. Given that the lengths of exponent and mantissa bits are the main distinctions among various \fp formats, it is reasonable to hypothesize that the proportion of exponent versus mantissa bits defined in each format can greatly influence the impact of a bit flip. Consequently, different \fp formats may exhibit diverse error resilience characteristics. Therefore, fault tolerance techniques should adopt precision-aware approaches when addressing applications that utilize mixed-precision \fp computations.

\subsection{The Example Platform: NVIDIA Tensor Core Architecture}

\begin{figure}[h!]
    \centering
    \includegraphics[width=0.9\linewidth]{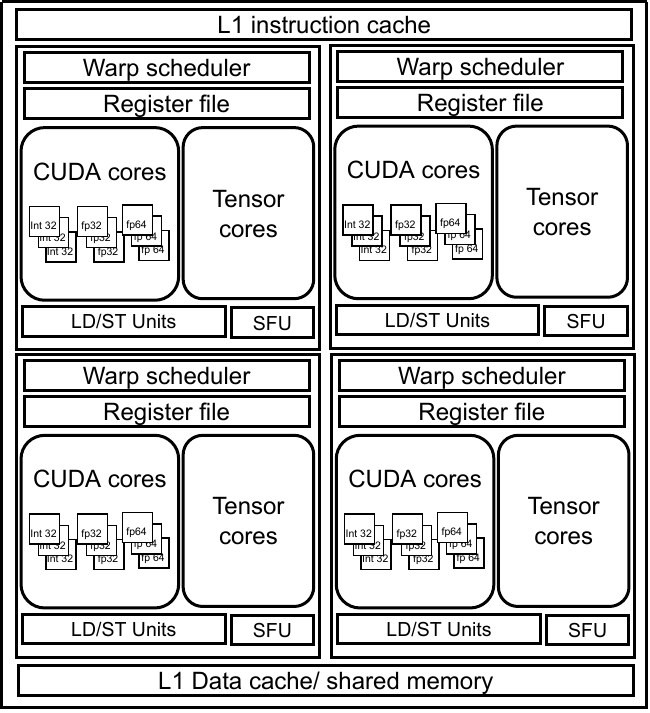}
    \caption{The simplified Ampere GPU SM architecture. Each SM contains 4 sets of CUDA cores and tensor cores. The CUDA cores and tensor cores share resources within the same SM.}
    \label{fig:tensorcode}
\end{figure}

Modern GPUs comprise multiple Streaming Multiprocessors (SMs) as the core computation units. Each SM in the NVIDIA Ampere architecture houses 4 Tensor Cores alongside the traditional CUDA cores, as shown in Figure~\ref{fig:tensorcode}. Tensor Cores are specifically designed to accelerate Matrix Multiplication (MM) operations, such as $D = A\times B + C$, where $A$ and $B$ are input matrices of shape $m \times k$ and $k \times n$ respectively, and the accumulating matrix $C$ with the shape $m \times n$.

The Ampere architecture introduces support for low-precision \bft and \tf formats in addition to \half, significantly enhancing the capabilities for deep learning applications. While previous works have extensively examined older generations of Tensor Cores, focusing on aspects like data movement, warp scheduling, and mixed-precision programming for MM~\cite{hongkong, tor, citidal}, the new architecture generation is beyond their scope. This study concentrates on the warp-level Half Matrix Multiply-Accumulate (\hmma) Source and Assembly (SASS) instructions, which form the core GEMM operations executed on Tensor Cores. An intriguing feature of the \hmma instructions is that they enable threads to share data by accessing other threads' registers within the same warp, a capability unavailable to regular CUDA Cores due to thread-independent registers.

Our research primarily centers around NVIDIA Tensor Core architectures, although our goals extend beyond a specific hardware platform. We chose the NVIDIA ecosystem for the following reasons: 1) The availability of toolchains from publicly-available resources defines the scope of our fault injection approaches tailored for mixed-precision GEMM computation; 2) The state-of-the-art machine learning infrastructure possesses the necessary capabilities to execute model inferences using various types of \fp formats. Importantly, the principles and techniques we employed can be applied to other hardware architectures if similar tools are available, i.e., a sophisticated and publicly available program instrumental framework and execution environment allowing for the analysis and instrumentation of mixed-precision GEMM operations.

\subsection{NVBitFI and \ours}

NVBitFI~\cite{nvbitfi} is a software-based fault injection framework built on NVBit~\cite{nvbit}, NVIDIA's binary instrumentation tool. NVBitFI utilizes NVBit's instrumentation interface to insert a callback function after each GPU instruction (SASS) or a selected group of instructions during compile time. The framework operates in two phases: profiling and injection. In the profiling phase, fault injection sites are generated, and in the injection phase, an instruction within the site is randomly chosen for fault injection.

In this paper, we present \ours as an independent and advanced iteration of NVBitFI. Notably, \ours introduces the capability to specifically inject faults into GEMM instructions. Section~\ref{sec:method} outlines in detail how \ours systematically introduces faults for mixed-precision GEMM. Moreover, the techniques proposed by \ours can be adapted to other architectures that involve GEMM instructions, expanding its applicability beyond its initial scope.

\begin{figure*}[h]
     \centering
     \includegraphics[width=\textwidth]{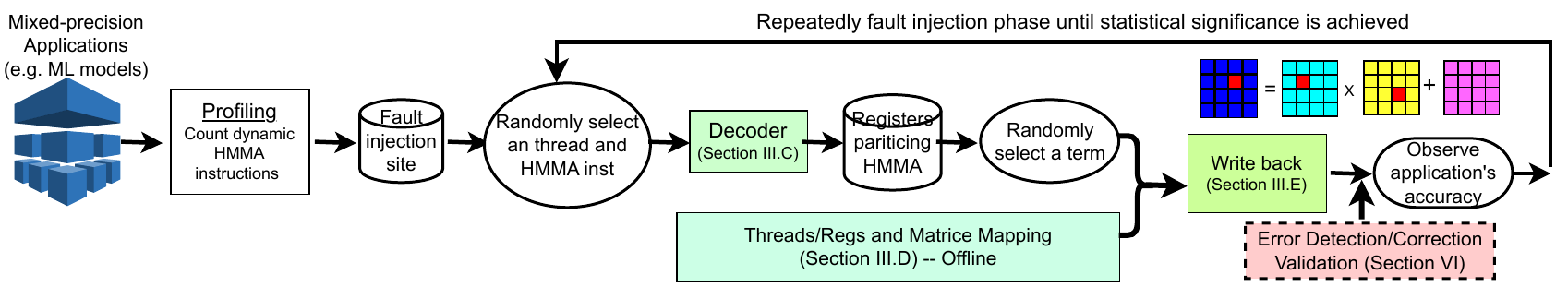}
     \caption{\ours's overall approach. \ours combines the offline (threads/registers and matrices mapping) and online modules to intercept a \hmma instruction execution and inject a bit-flip fault into the partial result of the dot-product calculation. The ``term" refers to a pair of matrix A and B elements participating in the dot-product. The error detection and correction validation (shown in the dotted box) indicates how we evaluate the protection mechanisms within the \ours framework.}
     \label{fig:overview}
     \vspace{-0.1in}
\end{figure*}

\section{\ours: GEMM enabled Fault Injection}
\label{sec:method}


This section presents the design and implementation of \ours. Specifically, we examine the matrix multiplication and addition operations of Tensor Cores' \hmma instructions. We observe how a \hmma instruction maps and operates on matrix data and then leverage these observations to facilitate fault injections in \ours.

\subsection{Challenges of Injecting Faults into \hmma Instructions}
  

\hmma instructions are highly intricate, involving multiple registers and threads across thread warps in every execution. One of the key features of \hmma instructions is the collective operations among the threads within a warp. A thread within the warp fetches matrix element data from other threads' registers, and the computation results contribute to four elements of the output matrix. This poses two challenges for \ours:

\noindent \textbf{1. Characterization}: To inject a fault into \hmma, \ours needs to intercept the \hmma instruction while mixed-precision operations occur during the dot-product calculation. Thus, a simple bit flip in the destination register of a \hmma instruction is not representative enough for characterization.


\noindent \textbf{2. Uniform Fault Injection Target Selection}: To ensure statistical significance, the fault injection process must uniformly select a thread combination and a \hmma instruction for each injection. However, the mapping details between the threads' registers and the matrix elements are not publicly exposed. Therefore, \ours must understand which elements of matrix $D$ the \hmma produces and which elements of matrices $A$ and $B$ get involved for a random \hmma instruction of a random GPU thread under different \fp formats.


\subsection{\ours Overview}

Figure~\ref{fig:overview} provides an overview of the \ours methodology. Since our study focuses solely on GEMM instructions, \ours generates the fault injection site (i.e., where the faults can be injected during the execution of the application) by counting only the dynamically executed \hmma instructions from all GPU threads. A random \hmma instruction is selected during each fault injection trial, and the GPU context switches to the thread executing this instruction to inject a fault. To identify which threads these two registers come from, \ours conducts the threads/registers and matrices mapping, which is an offline module (Section~\ref{sec:method_map}) inferring how Tensor Cores distribute the matrices within the warp and the exact mapping between matrix elements and the local registers of the threads in the warp. A dedicated \hmma decoder of \ours takes the selected \hmma instruction as input and outputs the register pairs, i.e., two registers, each holding one element of matrix A and B, respectively (Section~\ref{sec:method_de}). Finally, the write-back module implements the actual fault injection functions by referring to the mapping module to obtain the corresponding values of the randomly selected register pair.

\subsection{Decoding an \hmma instruction}
\label{sec:method_de}

\begin{figure}[t]
    \centering
    \includegraphics[width=0.35\textwidth]{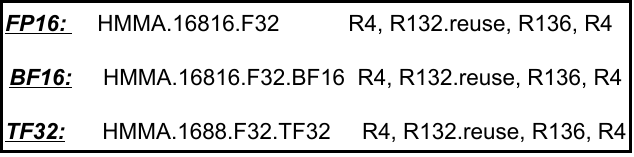}
    \caption{Three example \hmma instructions for GEMM  operations. Each \fp format generates its own \hmma instructions.}
    \label{fig:hmma_inst}
    \vspace{-0.2in}
\end{figure}

Over the evolution of the Tensor Cores and CUDA SDKs, \hmma instructions have been redesigned to leverage the capabilities of new architectures. A prior study~\cite{dissect_ang} provides a detailed analysis of how a matrix-multiply-and-accumulate (MMA) operation is compiled into one or a set of \hmma instructions for Volta, Turing, and Ampere Tensor Core architectures. In our study, we focus on the \hmma instruction generated for Ampere Tensor Cores and highlight the differences in \hmma instructions for different \fp formats. Figure~\ref{fig:hmma_inst} displays the \hmma instructions generated for \half, \bft, and \tf formats.\footnote{The example is compiled with SM80 and CUDA 11.6.} A \hmma instruction consists of three components: the opcode, the destination register(s), and the operands. Note that for \hmma instructions, some registers implicitly participate in the computation but are not shown in SASS.


\begin{itemize}[topsep=0pt,leftmargin=*]
\renewcommand{\labelitemi}{\scriptsize$\blacksquare$}
\item \textit{Opcode:} The \texttt{.16816} and \texttt{.1688} infixes after the "HMMA" indicate the shape of the matrix multiplication ($16\times 8\times 16$ for \half and \bft, and $16\times 8\times 8$ for \tf). The infix \texttt{.F32} after the shape indicator denotes the accumulate type, while the infix \texttt{.BF16} or \texttt{.TF32} indicates the matrix multiply type.
\item \textit{Destination Register(s):} The \texttt{R4} register is the explicit operand for the destination register. Implicitly, three more destination registers (\texttt{R5, R6, R7}) are associated with this \hmma instruction, each accumulating the computation results, respectively.
\item \textit{Operands:} For the input matrix $A$, the registers \texttt{R132} (explicitly) and \texttt{R133, R134, and R135} (implicitly) hold the elements. For the input matrix $B$, only the registers \texttt{R136} (explicitly) and \texttt{R137} (implicitly) are used. The register \texttt{R4} in the last part of the instruction stores the elements of matrix $C$ as the accumulator. For \half and \bft precision, since each GPU register occupies 32 bits, two \half or \bft values can be stored in one register, whereas only one \tf value can be stored in a register, as the \tf format requires 19 bits.
\end{itemize}

\begin{figure*}[h]
     \centering
     \includegraphics[width=0.85\textwidth]{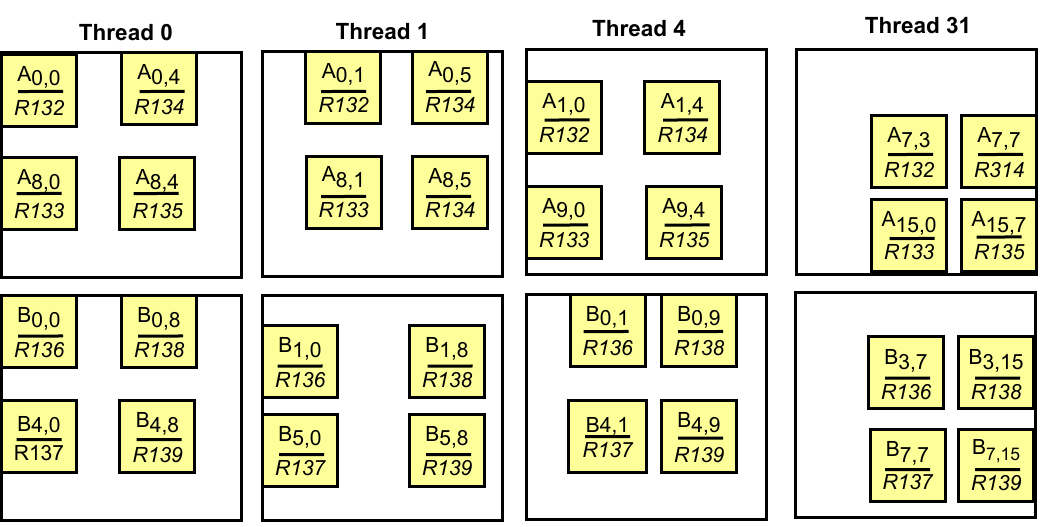}
     \caption{The registers in threads 0, 1, 4, and 31 and the elements of the $16\times 8$ portion of $A$ and $B$ stored in the registers. The indices are with respect to the column-major layout. Note the position of an element is disproportional in the matrix layout}
     \label{fig:layout}
     \vspace{-0.1in}
\end{figure*}

\noindent \textbf{Thread/Register Buffer:} During the fault injection phase, \ours captures the values of registers associated with the selected \hmma instruction from all threads within the same warp as the executing GPU thread (\textit{InjectedThread}). This process creates a lookup table that enables \textit{InjectedThread} to access the values of registers designated for fault injection directly. By maintaining this thread/register buffer, \ours ensures precise and consistent fault introduction, allowing for controlled and targeted fault injection.


\subsection{Threads/register and matrix data elements mapping}
\label{sec:method_map}



One of the main challenges in conducting fault injections for GEMM instructions is understanding how matrices $A$ and $B$ are mapped across threads and how different threads contribute to the GEMM computation during the execution of an \hmma instruction. To address this challenge, we utilize a GEMM benchmark from the NVIDIA CUTLASS library~\cite{cutlass} and employ CUDA-GDB~\cite{cudagdb} to track the register values of every thread in a warp. This mapping varies significantly based on the Tensor Core generations and the \fp formats. In this study, we investigate the internal mechanisms of Ampere Tensor Cores and present our observations below.

\begin{figure}[h!]
    \vspace{-0.1in}
     \begin{subfigure}[t]{0.15\textwidth}
         \centering
         \includegraphics[width=\textwidth]{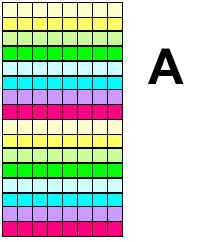}
         \label{fig:ma_tg}
     \end{subfigure}
          \hfill
     \begin{subfigure}[t]{0.26\textwidth}
       \centering
         \includegraphics[width=\textwidth]{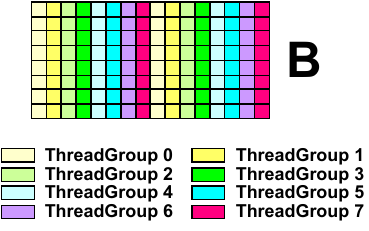}
         \label{fig:mb_tg}
     \end{subfigure}
    \vspace{-1em}
     \caption{The element layout for a $16\times8$ region of $A$ and a $8\times16$ region of $B$. The ThreadGroup ID = $\left \lfloor \frac{Thread ID}{4} \right \rfloor$}
     \label{fig:tg_view}
     \vspace{-0.05in}
\end{figure}

\noindent \textbf{1. Register Storage Layout:} Figure~\ref{fig:layout} shows how the elements of matrices $A$ and $B$ are stored in the local registers of threads when using the \tf data type. Notably, both $A$ and $B$ are stored in column-major order. The example showcases the column-major indices for threads 0 to 3, where their $R132$s and $R134$s hold the first 8 elements of row 0 in matrix $A$, while their $R133$s and $R135$s store the first 8 elements of row 8 in $A$. Similarly, threads 4 to 7 handle the first 8 elements of rows 1 and 9 in matrix $A$, respectively. For matrix $B$, the element storage is vertical. Threads 0 to 3 and 4 to 7 form two separate \textit{ThreadGroups}, a concept introduced in Tensor Cores to organize four consecutive threads in a warp. The example illustrated in Figure~\ref{fig:layout} presents the data storage layout of four threads: the first two threads in \textit{ThreadGroup} 0 (i.e., thread 0 and 1 as the thread ID offset of a warp), the first thread in \textit{ThreadGroup} 1 (i.e., thread 4 globally in a warp), and the last thread in the last \textit{ThreadGroup} 7 of a warp (i.e., thread 31 of a warp). Figure~\ref{fig:tg_view} provides the complete element storage layout with the view of the \textit{ThreadGroup} for matrices $A$ and $B$.

\begin{figure*}[h]
       \centering
         \includegraphics[width=0.9\textwidth]{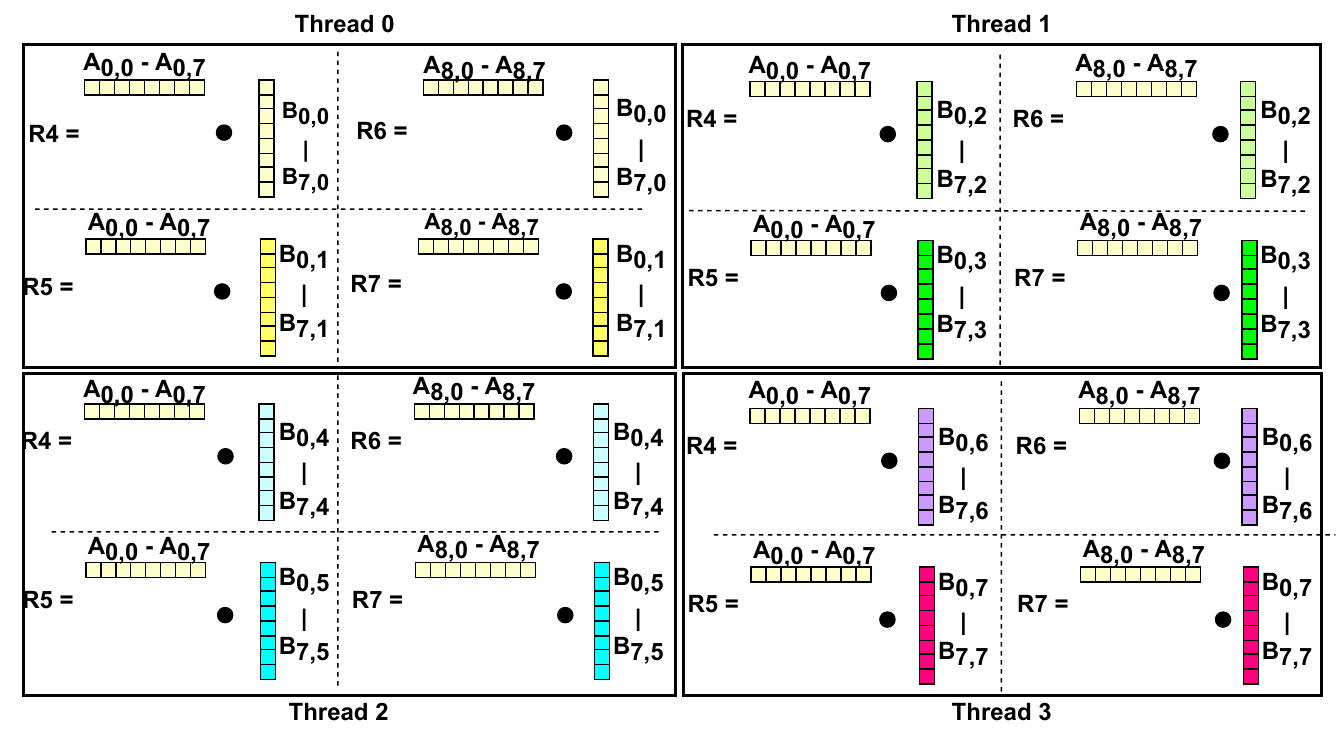}
         \caption{The pattern for the example \hmma instruction (Figure~\ref{fig:hmma_inst}, \tf format) to compute the dot products. Each thread within \textit{ThreadGroup 0} performs 4 dot product computations and stores the result in resulting registers. The rest of the \textit{ThreadGroups} follow the same pattern by taking their corresponding data from the $A$ and $B$ as the \textit{ThreadGroup 0}.}
         \label{fig:hmma_compute}
         \vspace{-0.1in}
\end{figure*}

\noindent \textbf{2. Computation Pattern:} The computation pattern for each thread within a \textit{ThreadGroup} depends on its relative thread ID within the warp. Figure~\ref{fig:hmma_compute} illustrates how \textit{ThreadGroup} 0 executes the \tf-based instructions shown in Figure~\ref{fig:hmma_inst}. For each destination register $R4, R5, R6, R7$, Figure~\ref{fig:hmma_compute} reveals the elements from matrix $A$ and matrix $B$ participating in the dot product. Specifically, the involved elements of matrix $A$ are either $A_{0,0} - A_{0,7}$ or $A_{7,0} - A_{7,7}$, while the \hmma instruction sweeps through the first $8\times 8$ elements of matrix $B$. This pattern remains consistent for the other \textit{ThreadGroups}, where each \textit{ThreadGroup} fetches the elements of matrix $A$ based on their storage layout shown in Figure~\ref{fig:tg_view} and performs the $8\times 8$ element sweep of matrix $B$ column by column, as indicated in Figure~\ref{fig:hmma_compute}. The overall process entails a $16\times 8 \times 8$ matrix multiplication computation.


The storage layout for the \bft and \half formats is consistent with \tf. Each 32-bit register (the size of a regular register in CUDA) can hold two 16-bit values. Thus, each register accommodates two consecutive elements of the matrix. In this scenario, 8 \textit{ThreadGroups} of a warp handle $16\times 16$ elements of matrix $A$ and $16\times 8$ elements of matrix $B$, resulting in a $16\times 8 \times 16$ matrix multiplication computation.

\subsection{Fault Injection with the Write-Back Approach}

\begin{table*}[tp]
\centering
\caption{DNN models used for error resilience characterization of mixed-precision \fp computation}
\begin{tabular}{llllll}
\hline
\multirow{2}{*}{\textbf{Model}} & \multirow{2}{*}{\textbf{Dataset}} & \multirow{2}{*}{\textbf{Description}}                                                                            & \multicolumn{3}{c}{\textbf{Test accuracy/reward}} \\ \cline{4-6} 
                       &                          &                                                                                                         & BF16         & FP16        & TF32        \\ \hline
ResNet-18              & CIFAR10                  & \begin{tabular}[c]{@{}l@{}}Deep residual network\\  for image classification\end{tabular}       & 93.0\%       & 92.2\%      & 92.1\%      \\ \hline
BERT                   & GLUE: MRPC               & \begin{tabular}[c]{@{}l@{}}Natural language modeling \\ with Transformer models\end{tabular}          & 81.6\%       & 82.0\%      & 82.0\%      \\ \hline
iGPT                   & Fashion MNIST              & \begin{tabular}[c]{@{}l@{}}Generative pre-training method \\ learning image representations\end{tabular} & 85.7\%       & 85.4\%      & 86.2\%      \\ \hline
UNet           & KITTI                    & \begin{tabular}[c]{@{}l@{}}Multi-class semantic segmentation \\ on images \end{tabular}                 & 84.9\%       & 84.9\%      & 85.1\%      \\ \hline
DQN                    & GYM: PongNoFrameskip-v4  & \begin{tabular}[c]{@{}l@{}}A deep Q-learning reinforcement\\ learning agent\end{tabular}     & 18.0         & 21.0        &    18.0         \\ \hline
\end{tabular}
\label{tab:benchmark}
\end{table*}

Upon selecting a random \hmma instruction and the corresponding executing thread, \ours utilizes the mapping information to identify the data source (i.e., other threads/registers from the warp). For instance, in the computation $\sum_{i=0}^{7}A_{0,i} * B_{i,0}$ shown in Figure~\ref{fig:hmma_compute} for thread 0 as the fault injection candidate, \ours randomly chooses a term (e.g., $A_{0,i} * B_{i,0}$ where $i\in [0,7]$), obtains the register values from the thread/register buffer, and flips a bit in the multiplication result.

To mimic the error's effect on the mixed-precision \fp computation, \ours employs a process called the ``write-back'' approach. It first stores the original value of the dot product as $re_{sum} = \sum_{i=0}^{7}A_{0, i} * B_{i,0}$. Then, \ours randomly selects a term, say $A_{0,2} * B_{2,0}$, and injects an error by flipping one of the bits in the result $re_{2}$, resulting in $re_{err}$. Next, \ours performs the following computation and stores the modified value ${re}'_{sum}$ back to $R4$:
\begin{align}
    {re}' = re_{sum} - re_{2} \\
    {re}'_{sum} = {re}' + re_{err}
\label{eq:sum}
\end{align}





\section{Experimental Methodology}
\label{sec:em}

\subsection{Fault Injection Methodology} 

We employ \ours to conduct large-scale fault injection campaigns on five representative DNN models. In each fault injection trial, \ours randomly selects a kernel, a thread in that kernel, a \hmma instruction executed by the thread, a random destination register of that \hmma instruction, and a random term contributing to the result. Note that for \tf, the total number of terms is 8, while for \bft and \half, it is 16. Upon selecting a term, \ours flips a random bit(s) in its resulting \fp value. The ``write-back'' approach is then applied to propagate the erroneous value to the destination register of the \hmma instruction. A single fault is injected in each fault injection trial.

\begin{figure*}
\captionsetup[subfigure]{labelformat=empty}
    \centering
    \subfloat[\centering ]
    {
    {\includegraphics[width=0.24\textwidth]{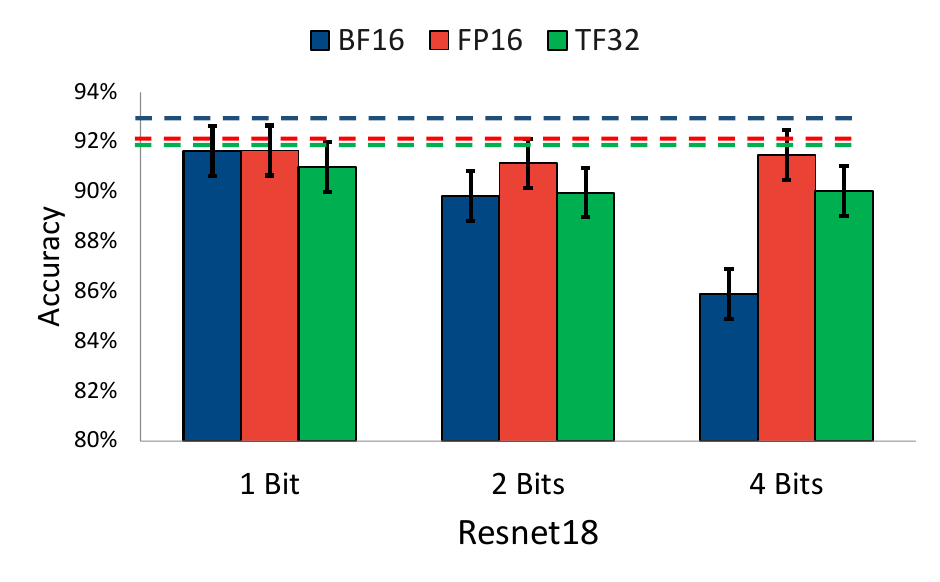} }
    \label{fig:resnet}%
    }
    \subfloat[\centering ]
    {
    {\includegraphics[width=0.24\textwidth]{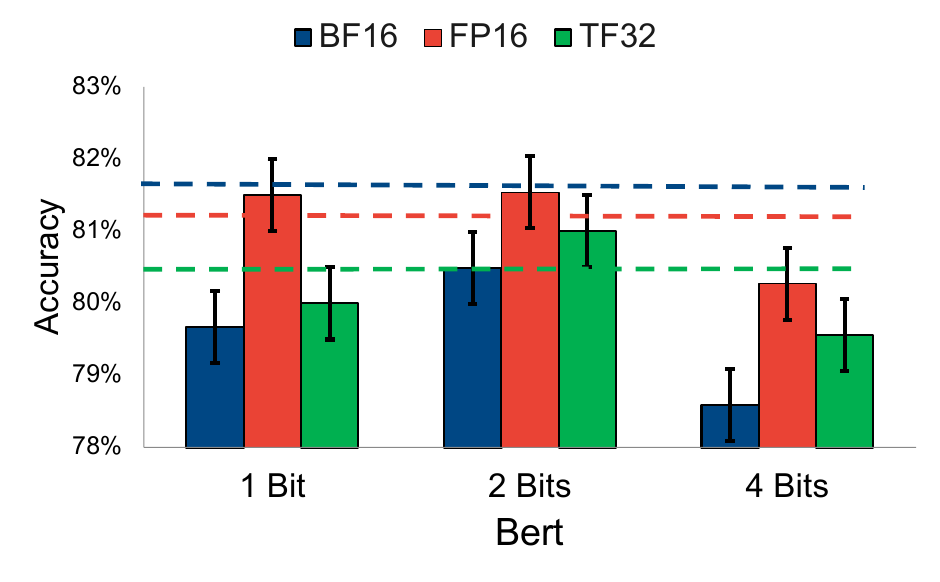} }
    \label{fig:bert}%
    }
    \subfloat[\centering ]
    {
    {\includegraphics[width=0.24\textwidth]{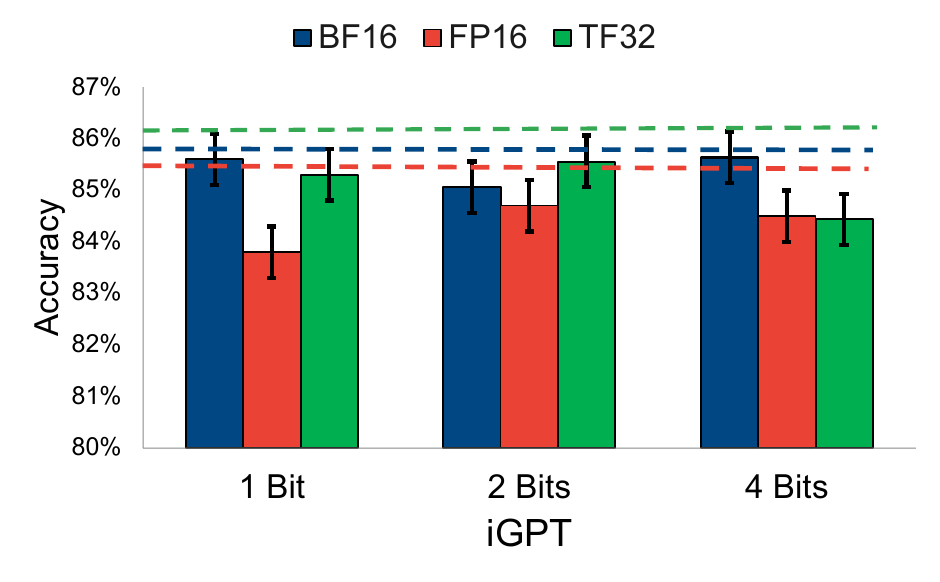} }
    \label{fig:igpt}%
    }
     \subfloat[\centering ]
    {{\includegraphics[width=0.24\textwidth]{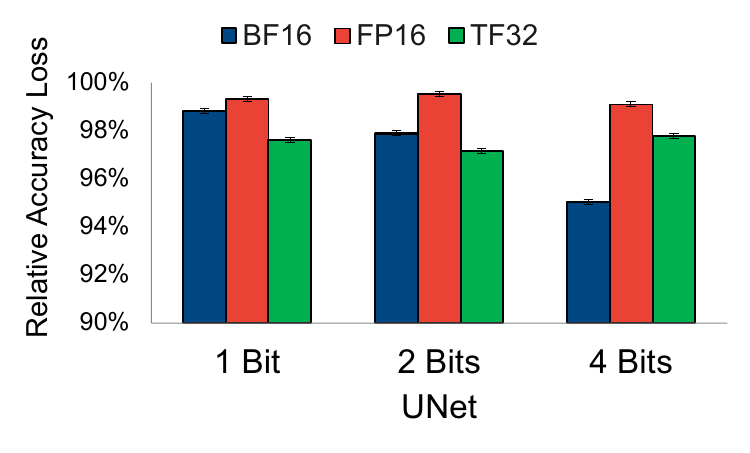}}
    \label{fig:seg}%
    }
   \vspace{-0.8cm}
    \caption{The fault injection results for four DNN models that exhibit observable accuracy loss across three \fp formats. The fault-free accuracy is shown as the dotted line for the reference matching with the \fp format color bar. The result for DQN is not shown, as the ratio of the number of cases where the average reward decreases is insignificant.}
    \label{fig:eval_results}
    \vspace{-0.1in}
\end{figure*}

\subsection{Benchmark models}

 

We instrument five DNN models to utilize NVIDIA tensor cores. These models are implemented using the PyTorch Lightning~\cite{pytorchlightning} framework, which facilitates convenient experimentation by encapsulating functional elements of a PyTorch DNN implementation. The Trainer class in PyTorch Lightning specifies the precision as "bf16" (i.e., \bft) or "16" (i.e., \half) to enable lower precision training and testing for the model. PyTorch also provides an interface to utilize \tf on NVIDIA Ampere tensor cores.

Table~\ref{tab:benchmark} describes the five representative models and the datasets used in our evaluation. It reports the fault-free test accuracy trained with \bft, \half, and \tf \fp formats for ResNet-18, BERT, iGPT, and UNet. For DQN, the converged reward values with each \fp format are reported instead.



\noindent \textbf{Estimating the Model Accuracy Degradation:} To assess whether a transient hardware fault impacts a model's performance, we employ three estimation methods tailored to the models' characteristics:

\noindent i). For ResNet-18, BERT, and iGPT models, a single sample is randomly selected from the test dataset in each fault injection trial. This emulates the actual inference task. Multiple fault injection trials yield a representative accuracy for the model.

\noindent ii). For the UNet model, a single sample is chosen from the test dataset in each fault injection trial. In this case, the model accuracy is determined by computing the ratio of correctly classified pixels to all pixels. The relative accuracy loss is reported for this model.

\noindent iii). For the DQN model, an average reward over test episodes is reported. We set the batch size to 1 during fault injection trials, leading to a constant average reward value for each fault-free run. Therefore, the relative reward loss is reported as the fault injection outcome.

\begin{table*}[]
\centering
\caption{Percentage of zero differences across bit flips and \fp formats for each model}
\begin{tabular}{lrrrrrrrrr}
\hline
\multirow{2}{*}{Model} & \multicolumn{3}{c}{BF16}                                                         & \multicolumn{3}{c}{FP16}                                     & \multicolumn{3}{c}{TF32}                                                         \\ \cline{2-10} 
                       & \multicolumn{1}{c}{1 Bit} & \multicolumn{1}{c}{2 Bits} & \multicolumn{1}{c}{4 Bits} & \multicolumn{1}{c}{1 Bit} & \multicolumn{1}{c}{2 Bits} & 4 Bits & \multicolumn{1}{c}{1 Bit} & \multicolumn{1}{c}{2 Bits} & \multicolumn{1}{c}{4 Bits} \\ \hline
ResNet-18              & 60.0\%                   & 57.0\%                    & 49.0\%                    & 24.0\%                   & 5.0\%                     & 0.4\% & 86.0\%                   & 83.0\%                    & 70.0\%                    \\ \hline
BERT                   & 3.0\%                    & 3.0\%                     & 2.0\%                     & 3.0\%                    & 0.3\%                     & 0.2\% & 10.0\%                   & 2.0\%                     & 2.0\%                     \\ \hline
iGPT                   & 89.0\%                   & 82.0\%                    & 71.0\%                    & 29.0\%                   & 5.0\%                     & 0.3\% & 84.0\%                   & 78.0\%                    & 68.0\%                    \\ \hline
UNet           & 67.0\%                   & 63.0\%                    & 52.0\%                    & 26.0\%                   & 5.0\%                     & 0.4\% & 71.0\%                   & 68.0\%                    & 69.0\%                    \\ \hline
DQN           & 63.0\%                   & 74.3\%                    & 30.0\%                    & 10.0\%                   & 9.0\%                     & 0.6\% & 94.0\%                   & 88.0\%                    & 77.0\%                    \\
\hline
\end{tabular}
\label{tab:no_diff}
\end{table*}

\subsection{Fault model}
We ensure the validity and reliability of our findings by meticulously designing and executing fault injection experiments. We assess the resilience of a model against errors in three \fp formats using single and multiple bit-flips as our fault model. The \hmma instruction involves multiplication operations between two lower-precision floating-point numbers and addition operations accumulated to a single \texttt{FP32} number. Hence, our study evaluates the impact of faults on the multiplication operations within the \hmma instruction. Specifically, we configure the fault injection experiment to flip 1 bit, 2 bits, or 4 bits within the result of a multiplication operation in each trial.

\begin{figure*}
    \centering
    \subfloat[\centering ResNet-18 \bft non-zero difference]
    {
    {\includegraphics[width=0.33\textwidth]{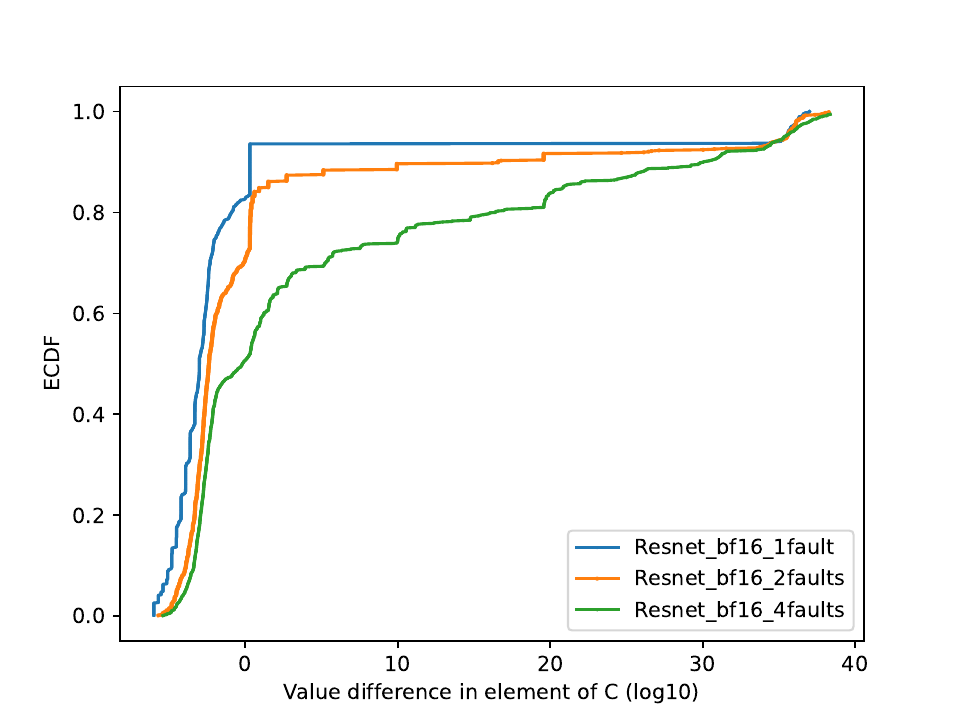} }
    \label{fig:resnet_bf_diff}%
    }
    \subfloat[\centering ResNet-18 \half non-zero difference]
    {
    {\includegraphics[width=0.33\textwidth]{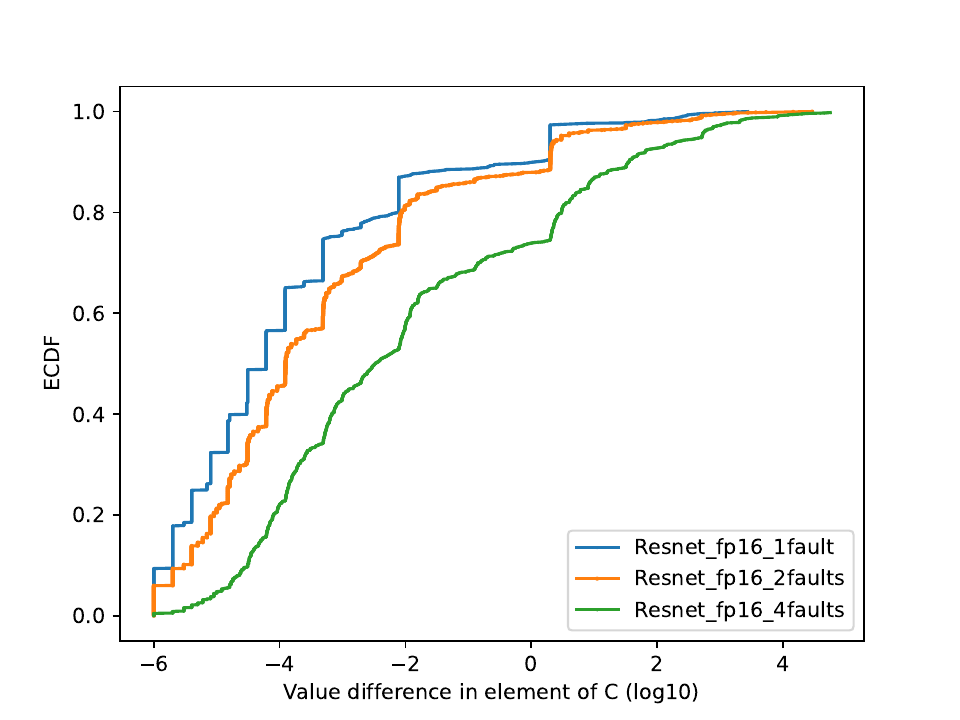} }
    \label{fig:resnet_fp_diff}%
    }
    \subfloat[\centering ResNet-18 \tf non-zero difference]
    {
    {\includegraphics[width=0.33\textwidth]{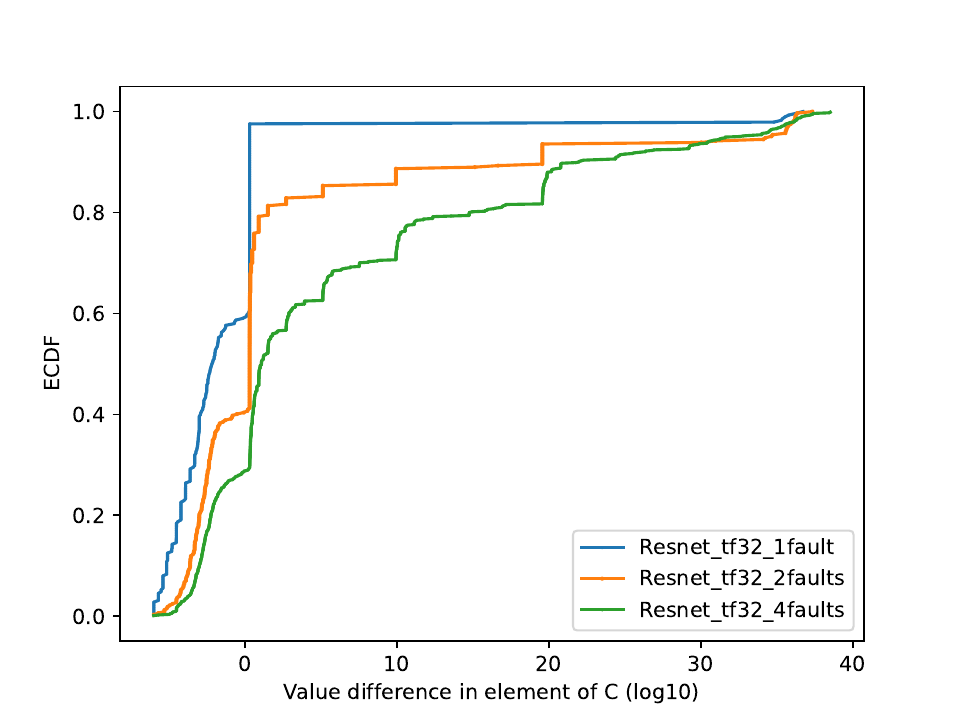} }
    \label{fig:resnet_tf_diff}%
    }
    
     \subfloat[\centering BERT \bft non-zero difference]
    {{\includegraphics[width=0.33\textwidth]{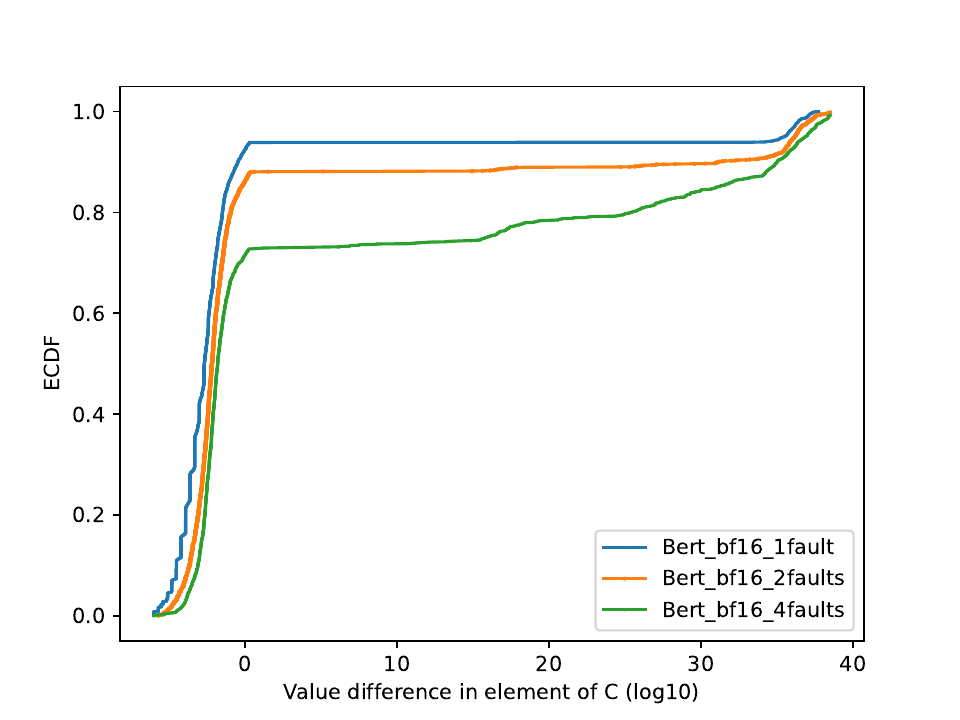}}
    \label{fig:seg}%
    }
     \subfloat[\centering BERT \half non-zero difference]
    {{\includegraphics[width=0.33\textwidth]{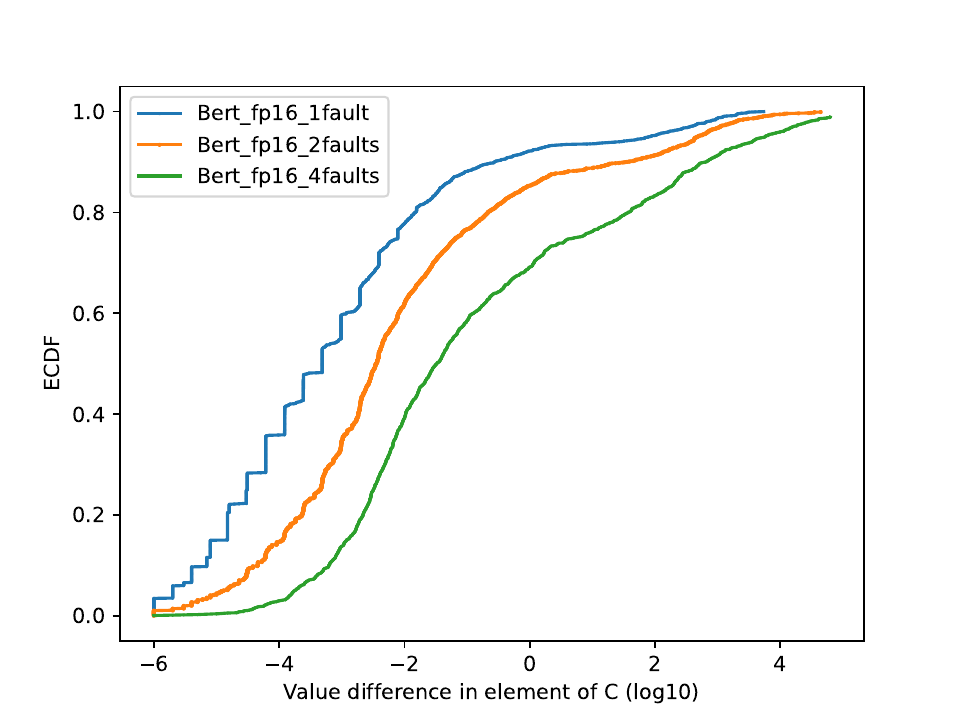}}
    \label{fig:seg}%
    }
     \subfloat[\centering BERT \tf non-zero difference]
    {{\includegraphics[width=0.33\textwidth]{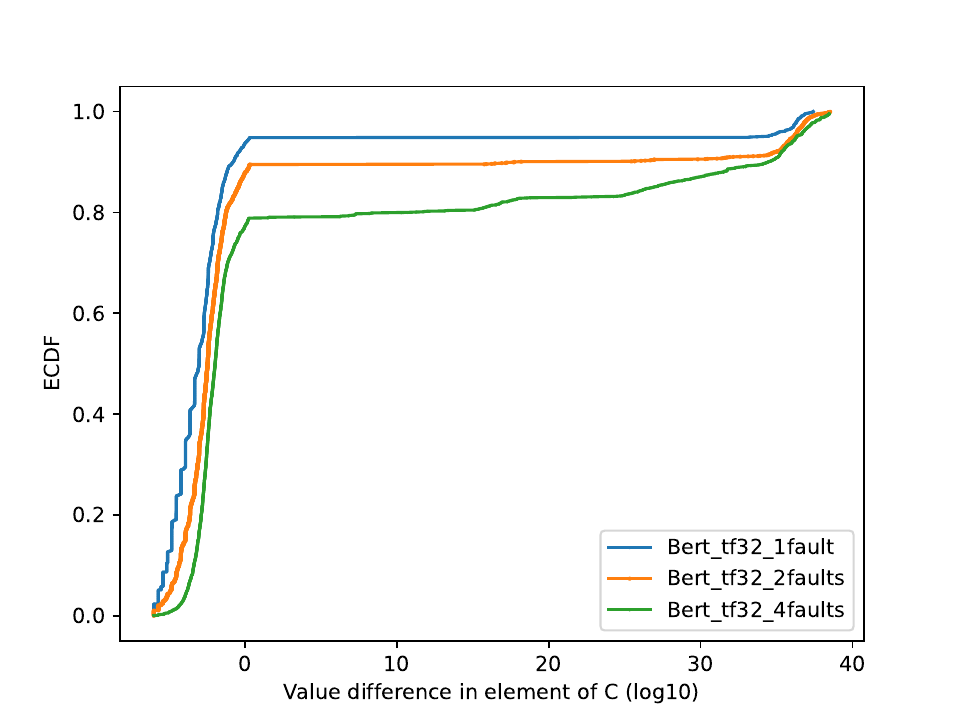}}
    \label{fig:}%
    }
    \caption{ECDF shows the distribution of the various differences for each configuration of the fault injection experiments on ResNet-18 and BERT. Other models exhibit similar distributions. The actual difference numbers are shown in the log scale.}
    \label{fig:diff}
    \vspace{-0.1in}
\end{figure*}


\section{Evaluation Results}
\label{sec:eval}

This section presents the result of error resilience characteristics of the common DNN models exploring mixed-precision GEMM. In addition to evaluating the model accuracy affected by the fault injection experiments, we also examine the values of original and injected cases to highlight the differences reflecting the numeric behaviours across different formats. 

Our hypothesis posits that the deviation of a faulty \fp number depends on the positions of the affected bits~\footnote{The sign bit is not discussed in this study as it consumes only one bit}. Flipping exponent bits, which determine the location of the decimal point, should have a greater impact on the resulting value than flipping mantissa bits, which hold the main digits. Therefore, we expect the largest deviation to occur in the \bft format: it has the highest ratio of exponent bits to mantissa bits compared to the \half and \tf formats.

\subsection{Overall injection results}
Figure~\ref{fig:eval_results} shows the accuracy of each model measured after injecting 1, 2, and 4-bit flips. Among the five models, DQN exhibits the most resilience to errors. The percentage of the affected number of relative rewards is less than 0.5\% across all 3 \fp formats under all configurations. Hence we do not present the results of DQN in Figure~\ref{fig:eval_results} due to the page limit.

A general trend observed across \bft, \half and \tf is that except for DQN, the rest of the models show less error resilience to faults when more bits are affected, and the decrease of the model accuracy becomes statistically significant for some models. Among the three \fp formats, \bft is the most vulnerable. The largest accuracy drop for \bft is from 93\% to 85.9\% (by 7\% in absolute accuracy) in ResNet-18 and 81.6\% to 78.6\% (by 3\% in absolute accuracy) in BERT, both with 4 bit-flips faults. In contrast, for \half and \tf formats, a marginal decrease is encountered, i.e. 1.06\% absolute accuracy drop on average). For UNet, the relative accuracy loss for \bft is 4.96\% compared to 0.9\% for \half and 2\% for \tf. In iGPT, the fault injection experiments incur a relatively small and stable impact on the overall model accuracy for all three formats. In summary, across five models, \bft is less resilient than \half and \tf by 3.97x and 3.11x on average. 

\subsection{Numeric deviation analysis}
We then investigate how the bitflip faults affect the numerical behaviors of mixed-precision \fp operations. In particular, we calculate the difference between the original sum ($re_{sum}$) and the sum obtained after fault injection (${re}'_{sum}$), as defined in Equation~\ref{eq:sum} (Section~\ref{sec:method}), for each fault injection trial. We then present the statistical distribution of this difference for various mixed-precision \fp formats.

In many cases, the fault injection did not alter the original value, resulting in a difference of 0. Two possible scenarios explain this outcome: 

\noindent (i) The original value is 0, which is common in deep learning models' activations. In this case, a small fraction of bits being flipped (e.g., 1, 2, and 4 bits) eventually result in extremely small numbers that are considered 0. 

\noindent (ii) Some flipped bits in the mantissa part lead to a small variation in the FP value, which is mitigated during accumulation. Table~\ref{tab:no_diff} reports the percentage of zero-difference cases over the total number of fault injections. Unsurprisingly, cases with 4-bit flips are more likely to make non-zero changes. 

We found that the \half format has a significantly higher chance of converting 0 to a non-zero value in the ResNet-18, iGPT, UNet and DQN models. We speculate that the reason behind this is that the base value of 0 in \half format is $2^{-14}$ and is $2^{-126}$ for \bft and \tf formats. It requires more bits to flip for \bft and \tf values to become non-zero. An example is that \texttt{0000001000000000} in \half (i.e. the first mantissa bit is 1) equals 0.00003052 while \texttt{0000000001000000} in \bft (i.e. also the first mantissa bit is 1) equals a very small number.  



We then exclude the zero-difference cases and compute the non-zero-difference cases' empirical cumulative distribution function (ECDF), shown in Figure~\ref{fig:diff}. It presents the ECDFs for ResNet-18 and BERT, where similar trends are observed in other models. The main takeaways are:
\begin{itemize}[topsep=0pt,leftmargin=*]
\renewcommand{\labelitemi}{\scriptsize$\blacksquare$}
    \item \bft and \tf produce a much larger range of errors that deviate from the original sum value, where most of the difference values from \half formats range in a significantly smaller bound. Since \bft and \tf have the same length of exponents bits, their erroneous ranges are similar. 
    \item Comparing \mbox{Figures \ref{fig:resnet_bf_diff} and \ref{fig:resnet_tf_diff}}, $\sim$80\% of the differences are approaching 20 (log scale) in \bft and 80\% of \tf cases are close to 13 (log scale), which explains why the \bft format corresponds to the largest accuracy drop in ResNet-18. Similar findings also exist in other models where \bft leads to the largest degradation of the model accuracy.
\end{itemize}

To summarize, we present our analysis of the numeric deviation in mixed-precision computation caused by fault injection. The values in the \half format have a higher probability of being converted to non-zero values upon the injection of bitflip faults, while \tf and \bft numbers have a lower chance of staying the same numerically. However, in cases where such changes occur for \bft and \tf formats, the differences between the original and faulty values are generally greater than what the \half format could generate.
\section{Hardware-aware Fault Tolerance Design}
\label{sec:detection_correction}

In the previous sections, we introduce \ours and evaluate the error resilience of deep learning applications that use mixed-precision computation. In this section, we present our proposed hardware-efficient solution for detecting and correcting transient hardware faults that may occur during the computation of reduced-precision multiplication, aiming to safeguard the overall mixed-precision computation path.

\subsection{Design guidelines}

Extensive research has been conducted on the topic of checking the correctness of floating-point computation~\cite{cherubin2020tools}. However, estimation of the accumulated errors used for this purpose may not be able to detect floating-point errors induced by hardware faults. In the context of ML, where tasks often require error-resilient and instantaneous results, fault tolerance techniques must be cost-effective.

Traditional approaches for error detection and correction are through dual modular redundancy (DMR) and triple modular redundancy (TMR), as a transient hardware fault is likely to occur once in a short period of time. DMR and TMR require significant extra computation, area and power resources, limiting their adoption in real-world ML tasks. Our proposed techniques are inspired by the insights obtained from our characterization study. Among \bft, \half and \tf formats, the hypothesis is verified that \bft would have a higher chance of being impacted by the faults since it has more exponent bits than \half and fewer mantissa bits than \tf. Since the exponent bits determine the magnitude of the number, they potentially carry more weight in the \fp value. Therefore, we assume that the bit flips in the mantissa bits will likely result in benign outcomes, and we propose an exponent-only detection and correction approach. This approach is specialized for \bft computation, as shown in the fault injection results that faults in \bft format could cause the largest accuracy drop.

Our exponent-only approach takes the register that stores the result of the \bft multiplication and operates on its exponent bits. It does not need to duplicate the \fp multiply-and-accumulate operations but adds a few operations within the original logic. We propose three techniques, namely BoundCheck, RangeCheck-max and RangeCheck-flip for the exponent-only approach.  


\noindent \textbf{\textit{Bound check:}} If the leftmost bit of the exponent bits is 1, it introduces a constraint on the position of the second leftmost 1, ensuring a reasonable value for the floating-point (FP) representation. Table~\ref{tab:bf_bound} presents the possible configurations of 1s in the different positions of the exponent bits when the most significant bit of the exponent bit (MSB, same below) is 1 and their corresponding values.

As shown in Table~\ref{tab:bf_bound}, in cases where the MSB and the fourth bit from the left of the exponent part are both 1 (i.e., \texttt{10010000}), the value of the floating-point number represented by \bft will not be less than 131,072. In machine learning (ML) workloads, it is generally rare to encounter a normalized partial weight/activation larger than such scale. 

To detect the above cases, we conservatively check any combination of the MSB and any position to the left of the fourth bit of the exponent part containing 1s. When such combinations are found, the correction task flips any 1s in the \texttt{eee} positions of the exponent \texttt{1eeexxxx} as MSB = 1 and all other \texttt{eee} should not contain 1s, where \texttt{x} can be either 0 or 1.
\begin{table}[]
\caption{Bounded values when the MSB = 1 and the second leftmost 1 flows in the exponent bits of \bft}
\centering
\begin{tabular}{lll}
\hline
Exponent (8bit)           & Mantissa (7bit) & Value       \\ \hline
\multirow{2}{*}{11000000} & 1111111         &  7.34987e+19 \\
                          & 0000000        & 3.68935e+19 \\ \hline
\multirow{2}{*}{10100000} & 1111111        & 1.71128e+10 \\
                          & 0000000        & 8.58993e+09 \\  \hline
\multirow{2}{*}{10010000} & 1111111        & 261120      \\
                          & 0000000        & 131072      \\ \hline
\multirow{2}{*}{10001000} & 1111111         & 1020        \\
                          &0000000         & 512         \\  \hline
\multirow{2}{*}{10000100} & 1111111         & 63.75       \\
                          & 0000000         & 32          \\ \hline
\multirow{2}{*}{10000010} & 1111111         & 15.9375     \\
                          & 0000000         & 8           \\ \hline
\multirow{2}{*}{10000001} & 1111111         & 7.96875     \\
                          & 0000000         & 4           \\ \hline
\multirow{2}{*}{10000000} & 1111111         & 3.98438     \\
                          & 0000000         & 2.0         \\ \hline
\end{tabular}
\label{tab:bf_bound}
\end{table}

\noindent \textbf{\textit{Range check - \textbf{flip} and \textbf{max}}}: since the multiply operation occurs between two \bft numbers, the result of the multiplication's exponent part should also be bounded by the input numbers. Below we briefly induce the range of the resulting exponent for a multiplication:


Consider two floating-point numbers $x_1$ and $x_2$:
\begin{align*}
x_1 = (-1)^{s_1}\cdot m_1 \cdot 2^{e_1},\quad x_2 = (-1)^{s_2}\cdot m_2 \cdot 2^{e_2}
\end{align*}

To determine the upper bound of the result of the multiplication operation, we can assume, without loss of generality, that $s_1$ and $s_2$ are both 0.

\noindent The result of multiplication $rm$ can be written as:
\begin{align*}
    rm = m_1 \cdot 2^{e_1} \cdot m_2 \cdot 2^{e_2} = m_1\cdot m_2 \cdot 2^{(e_1+e_2)}
\end{align*}
Since $m_1\cdot m_2 < 4$, we can determine a tight upper bound for the exponent of the result $e_m$ as:
\begin{align}
e_m \leq e_1+e_2+1 
\end{align}

That said, the upper bound for the exponent in binary representation is $e_1+e_2+1$. Therefore, the error detection is conducted by checking if  $e_m^\prime$(i.e. the exponent of the erroneous result) is larger than $e_1 + e_2 + 1$. If detected, the correction can be either 
i). replace the $e_m^\prime$ with $e_1 + e_2 + 1$, or 
ii). flip the 1s from the rightmost bit position of the $e_m^\prime$, and stop when the resulting value of the $e_m^\prime$ smaller or equal to the upper bound\footnote{We omit the bias in the above range inference since the bias contributes only to the final representation of a floating-point number. Our range check approach is conducted correctly considering the effect of the bias.}. We refer to the former technique as \textit{RangeCheck-max} and the latter as \textit{RangeCheck-flip}. 

\subsection{Evaluation on ML applications}

To evaluate the performance of the proposed techniques, we use \ours to inject 4-bit faults for ResNet-18, BERT and UNet, and apply the techniques separately after the fault is injected to detect and correct the faulty value. The correction will only be triggered if the detection criteria are violated in the software implementation. Since with 4-bit faults, these models exhibit the largest accuracy loss with \bft. We measure the improvement the techniques achieve in reducing the accuracy loss of \bft computation. 

\begin{figure}[h!]
\vspace{-0.1in}
\centering
\includegraphics[width=0.48\textwidth]{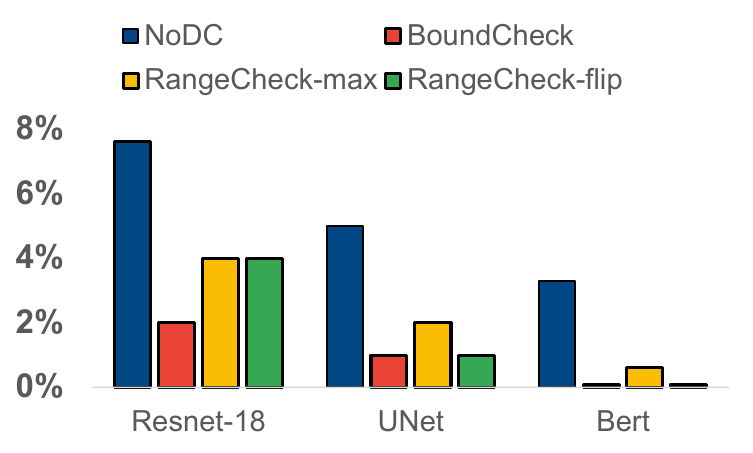}
\caption{The reduced accuracy loss after applying exponent-only techniques. NoDC represents the fault injection results without any detection and correction techniques. }
\label{fig:loss}
\vspace{-0.05in}
\end{figure}

Figure~\ref{fig:loss} shows the reduced accuracy loss for the three models after applying the proposed detection and correction techniques. For ResNet-18, the relative accuracy loss of 7.6\% (93\% to 85.9\%) decreases to 2\%, which translates to a 74\% improvement. For UNet, 60\% to 80\% improvement is obtained. The model achieves almost the same accuracy for BERT as the fault-free case. On average, 75\% improvement is achieved by applying our exponent-only techniques. 

\begin{figure}[h!]
\vspace{-0.2in}
\centering
\includegraphics[width=0.48\textwidth]{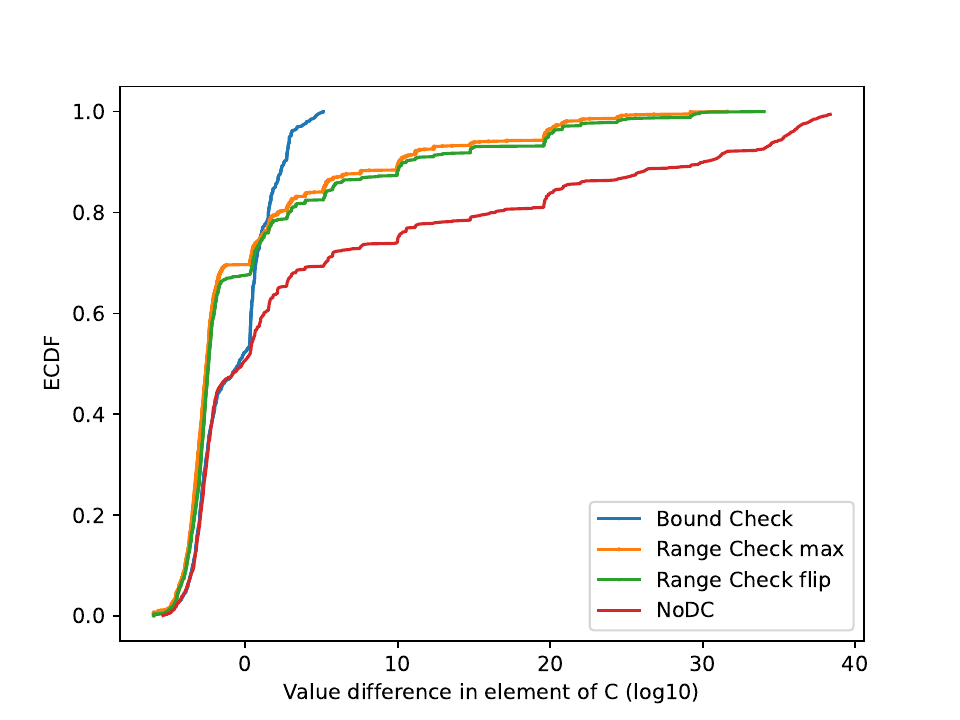}
\caption{For ResNet-18. The difference between the original and injected values with the correction technique applied. NoDC is the same as the green line in Figure~\ref{fig:resnet_bf_diff}. This explains why correction techniques work numerically.}
\label{fig:detect}
\vspace{-0.3cm}
\end{figure}

To further investigate how each technique affects the values before and after the fault injection, Figure~\ref{fig:detect} shows the ECDF of the difference for each technique and the original difference for ResNet-18. The difference generated with three correction techniques reaches 100\% much faster than that of NoDC, indicating that the corrected values are well confined in a reasonable range for ResNet-18. Figure~\ref{fig:detect} also helps understand why BoundCheck outperforms the other two techniques in improving the model accuracy. Due to the page limit, we didn't show such an analysis for UNet and Bert, yet both conclusions are consistent. 

\subsection{Hardware cost extrapolation}

\begin{figure}
    \centering
    \subfloat[\centering Bound check technique]
    {
    {\includegraphics[width=0.47\textwidth]{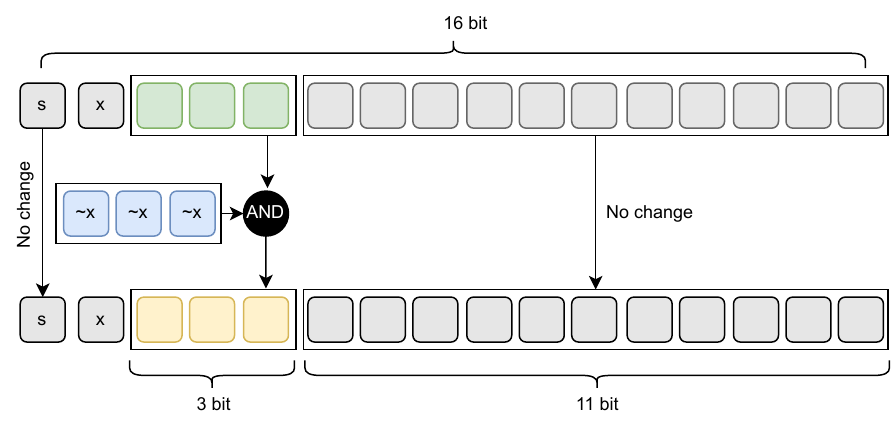} }
    \vspace{-0.2cm}
    \label{fig:boundcheck}%
    }\\
    \subfloat[\centering Range check with maximum possible exponent technique]
    {
    {\includegraphics[width=0.47\textwidth]{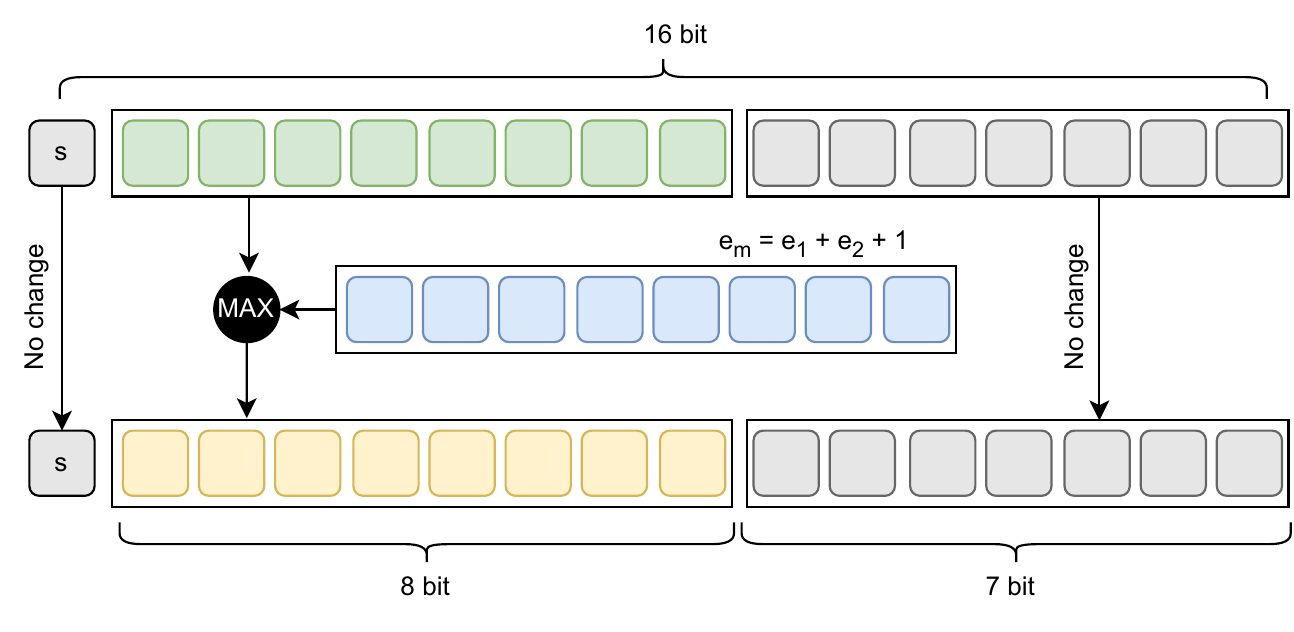} }
    \vspace{-0.2cm}
    \label{fig:rangecheck}%
    \vspace{-0.2cm}
    }
   
    \caption{Hardware design extrapolation for BoundCheck and RangeCheck-max techniques. RangeCheck-flip cannot be efficiently extrapolated due to the lack of implementation details.}
    \label{fig:hardware}
    \vspace{-0.4cm}
\end{figure}

This section presents our extrapolated hardware-oriented implementation of the proposed techniques. Figure~\ref{fig:hardware} depicts two possible designs of the proposed techniques. Note that to implement the RangeCheck-flip technique in hardware efficiently, one needs to understand the actual hardware design, like a detailed per-stage pipeline description. Therefore, we only present potential designs for BoundCheck and RangeCheck-max. Both designs only require logic and addition operations. Please note that the detection task is merged into the correction task to simplify the logic further. The BoundCheck in Figure~\ref{fig:boundcheck} takes three bit-wise inversions on the leftmost bit of the exponent and an \texttt{AND} operator to flip the out-of-bound exponent bits. Figure~\ref{fig:rangecheck} requires an addition to compute the maximum range for the new exponent and a \texttt{MAX} operation to choose the smaller exponent value for the result. Although the cost of a \texttt{MAX} depends on various factors, our design only performs the 8-bit comparison, which is the core operator inside the \texttt{MAX}.

When using the NVIDIA Tensor Core platform as an example, the multiply-accumulate (MAC) operation is performed in a multi-stage pipeline~\cite{nvidiamac}. The multiplier stage of this pipeline is the dominant factor in determining the latency of the operation. The critical path delay for floating-point multiplication can be up to seven times longer than for logical and addition operations, depending on the specific implementation of the multiplier~\cite{mac, mac2}. To minimize any additional overhead to the overall latency, the logical operations required by the three approaches can be synthesized into the stage immediately following the multiplier stage in the original Tensor Core pipeline. This arrangement ensures that the critical path remains dominated by the multiplier stage.
\section{Related work}

A few existing studies aim for DNN robustness against parameter perturbations or hardware faults. The traditional approach uses redundant hardware operating in parallel. For safety-critical applications such as in vehicles, triple modular redundancy has been used \cite{tmrcars}, but this has the disadvantages of triple the power consumption and chip area compared to conventional processors. Certain efforts select only the most sensitive computations to be duplicated \cite{selectiveprotection}, but this incurs doubling the cost for the selected computations.

Another branch focuses on robust training practices. Common regularization techniques include adding loss for high-magnitude parameters or using Dropout \cite{dropout} to make parameters less sensitive to bit flips. Other work on adversarial DNN \textit{inputs} \cite{adversarialexamples} shows that bit flips can be introduced during the training \cite{superoldfaulttoleranttraining} phase to increase the resilience of DNNs to bit flips, where such bit flips can come from \cite{biterrorrobustness} insufficient memory voltage \cite{undervoltingbitflip} or a targeted attack \cite{deephammer}. While regularization techniques are low-cost, they suffer from inefficiency \cite{biterrorrobustness}. Also, adversarial training or retraining \cite{oldfaulttoleranceretraining} takes additional training time, and all such techniques do not detect or correct errors that occur during inference.

Some work focused on modifying a trained DNN for better robustness. As DNN parameters are usually low magnitude, quantization to fixed-point numbers has been proposed to reduce sensitivity to bit flips \cite{biterrorrobustness, layerwisequantization}. To limit the impact of bit-flip errors, activation functions may be changed from (usually) $\mathrm{ReLU}$ to $\mathrm{ReLU6}$\footnote{$\mathrm{ReLU6}(x) = \max(0, \min(6, x))$} or $\tanh$ \cite{terminalbraindamage}, or range restrictions may be applied to their outputs \cite{ranger}. While such activation clamping has effectively increased overall resilience, it requires modification of the trained network.

There has been prior work on error detection and correction during inference. Because the least significant bits (LSBs) of DNN parameters do not affect accuracy much, \cite{opportunisticparity} flip the necessary LSBs such that the parity of all weights is even. If an odd number of bit flips occurred, they zero the weight. A limitation of this technique is that bit flips in activations cannot be addressed. One method of addressing activation errors takes advantage of the expectation that adjacent activations tend to have similar magnitude, \cite{justsayzero} zeros activations that deviate too much from adjacent activations in magnitude. A similar technique is zeroing layer outputs if they are statistically dissimilar to adjacent outputs \cite{neurongradientstatistics}, achieving state-of-the-art resilience on image classification tasks. A common element among these studies is zeroing anomalous values, which is reasonable given that DNN parameters are usually close to zero. However, it may still be preferable to maintain some of the clean parameters. Also, some of these detection methods incur computational costs, such as summing activations in the case of \cite{justsayzero} or statistical t-tests in the case of \cite{neurongradientstatistics}.


Thus, we highlight the advantages of our error detection and mitigation method: our techniques target mixed precision DNNs, and unlike prior work that requires either profiling the data distribution of the models or relying on model retraining to activate the available range checking, our proposed detection and correction techniques make no assumption on the model data nor incur training overhead. 


\section{Conclusion}
\label{sec:conclusion}
In this paper, we thoroughly investigate the impact of transient hardware faults on the error resilience of mixed-precision computation. To precisely target mixed-precision enabled GEMM instructions, we design and implement a fault injection tool named \ours. This tool effectively intercepts the \hmma instruction and introduces faults to its computation path. Leveraging \ours, we conduct extensive error resilience characterization on five representative machine learning models, gaining unique insights into the effects of different \fp formats. Among the three \fp formats considered, our findings reveal that \bft appears to be more vulnerable to hardware faults compared to \half and \tf when used in machine learning inference tasks. Building on these insights, we propose three lightweight error detection and correction techniques aimed at enhancing the overall error resilience of the models. 




\bibliographystyle{ieeetr}
\bibliography{citations}


\end{document}